# Organic Superalkalis with Closed-Shell Structure and Aromaticity

Ambrish Kumar Srivastava

*P. G. Department of Physics, Veer Kunwar Singh University, Ara- 802301, India*

E-mail: ambrishphysics@gmail.com

**Abstract**

Benzene ($C_6H_6$) and polycyclic hydrocarbons such as naphthalene ($C_{10}H_8$), anthracene ($C_{14}H_{10}$) and coronene ($C_{24}H_{12}$) are well known aromatic organic compounds. We study the substitution of Li replacing all H atoms in these hydrocarbons using density functional method. The vertical ionization energy (VIE) of such lithiated species, i.e., $C_6Li_6$, $C_{10}Li_8$, $C_{14}Li_{10}$ and $C_{24}Li_{12}$ ranges 4.24-4.50 eV, which is lower than the IE of Li atom. Thus, these species may behave as superalkalis, due to their lower IE than alkali metal. However, these lithiated species possess planar and closed-shell structure, unlike typical superalkalis. Furthermore, all Li-substituted species are aromatic although their degree of aromaticity is reduced as compared to corresponding hydrocarbon analogues. We have further explored the structure of $C_6Li_6$ as star-like, unlike its inorganic analogue $B_3N_3Li_6$, which appears as fan-like structure. We have also demonstrated that the interaction of $C_6Li_6$ with a superhalogen (such as $BF_4$) is similar to that of a typical superalkali (such as $OLi_3$). This may further suggest that the proposed lithiated species may form a new class of closed-shell organic superalkalis with aromaticity.

## 1. Introduction

Benzene ($C_6H_6$), a prototype six-membered ring, is the building block of aromatic organic species. Aromaticity is a distinguished feature of species, which leads to energy reduction and a variety of unusual chemical and physical properties [1] such as bond-length equalization, unusual reactivity, characteristic spectroscopic features etc. Aromatic rings are significantly stabilized, due to complete delocalization of $\pi$-electrons, which reduces their reactivity toward addition reactions. For instance, benzene possesses very large ionization energy and no positive electron affinity. This is why the whole chemistry of benzene is based on substitution reaction in which an atom or group usually replaces hydrogen attached to the ring. The substitution causes to significantly affect the chemical reactivity of the ring as noticed in many studies [2-4]. This may results in reducing the ionization energy or increasing the electron affinity of the substituted rings. Here, we study the substitution of lithium (Li) on benzene and its higher analogues such as naphthalene ($C_{10}H_8$), anthracene ($C_{14}H_{10}$) and coronene ($C_{24}H_{12}$) replacing all H-atoms. Few polylithium organic compounds have already been synthesized experimentally including $C_6Li_6$ [5, 6].

The species with lower IE than alkali atoms are referred to as superalkalis. According to Gutsev and Boldyrev [7, 8], such species can be designed by central electronegative atom with excess electropositive ligands. $FLi_2$, $OLi_3$, $NLi_4$ etc. are typical examples of superalkalis. These are hypervalent clusters possessing an excess electron and hence, open-shell structure. Therefore, they possess strong reducing capability and can be employed in the formation of a variety of charge transfer species with unusual properties. For instance, the use of superalkalis in the design of superbases with strong basicity [9-11] and supersalts with tailored properties [13-15] has been extensively studied. Owing to interesting properties of superalkalis and their compounds, such species have been continuously explored [16-21]. In this paper, we show that the vertical ionization energies (VIEs) of lithiated benzene ($C_6Li_6$) and its higher analogues such as $C_{10}Li_8$, $C_{14}Li_{10}$ and $C_{24}Li_{12}$ are lower than that of Li atom. Although, the aromaticity of the rings is affected by substitution of Li-atoms, their planarity is retained. We have also studied the interaction of $C_6Li_6$ with $BF_4$ superhalogen and compared with that of $OLi_3$ superalkali with $BF_4$. Note that $C_6Li_6$ has been previously studied by several groups [22-25]. The application of $C_6Li_6$ in the hydrogen storage has been also reported [26, 27]. We still believe that the low ionization energy feature of $C_6Li_6$ is probably studied here for the first time.

## 2. Computational methods

All ring structures considered in this study were fully optimized at B3LYP method [28, 29] using 6-311++G(d,p) basis set in Gaussian 09 program [30]. The geometry optimization was performed without any symmetry constraints and followed by frequency calculations to ensure that the optimized structures correspond to true minima in the potential energy surface. Our B3LYP computed C−C bond lengths (1.395 Å) and VIE (9.28 eV) of benzene are in good agreement with the experimental bond lengths of 1.399 Å measured by infrared spectroscopy [31] and ionization energy of 9.24 eV [32], respectively. On the contrary, the computed VIE of benzene at MP2 level is found to be 9.62 eV, which is overestimated as compared to B3LYP as well as experimental values. Therefore, we have adopted the present computational scheme. This results obtained by B3LYP scheme are easily reproducible at reduced computational cost, especially for larger systems such as $C_{24}H_{12}$, $C_{24}Li_{12}$, etc. where MP2 calculations become quite expensive.

The aromaticity of the ring structures are evaluated using various criteria such as nucleus independent chemical shifts (NICS and $NICS_{zz}$), harmonic oscillator model of aromaticity (HOMA) as well as para-delocalization index (PDI). NICS is the most popular magnetic criteria of aromaticity [33, 34], which has been calculated at the geometrical centre of the rings. HOMA index [35, 36] is a structural criteria, which quantifies the bond-length equalization of the rings. PDI [37], an electronic criteria is applicable six-membered ring systems only, which measures the delocalization of electrons within the ring.

## 3. Results and discussion

We start our discussion by considering benzene ($C_6H_6$), naphthalene ($C_{10}H_8$), anthracene ($C_{14}H_{10}$) and coronene ($C_{24}H_{12}$) as displayed in Fig. 1. Benzene is six-membered ring system with the C−C bond lengths of 1.395 Å. Naphthalene and anthracene consist of two and three fused benzene rings along an axis with the C−C bond-lengths of 1.374−1.420 Å and 1.367−1.443 Å, respectively. Coronene, on the contrary, contains seven fused rings in a plane such that the C−C bond-lengths become 1.370−1.426 Å, ranging between those of naphthalene and anthracene. All these ring structures are planar with C−H bond length of 1.085 Å. Next, the H-atoms of all these rings are substituted with Li-atoms. The resulting optimized structures are also displayed in Fig. 1 and corresponding parameters are listed in Table 1. Although all substituted ring structures retain their planarity like their hydrocarbon analogues, substituted Li-atoms bind with two carbons forming planar star-like structures. The tendency of Li atoms to bind with two C atoms has already been noticed in case of $C_6Li_6$ [22, 23].

The vertical ionization energy (VIE) of systems has been calculated by difference of total energy of equilibrium neutral structure and single-point energy of corresponding cation at the optimized geometry of neutral structure:

$$\text{VIE} = E_\text{cation} - E_\text{neutral}$$

The VIE of hydrocarbons considered ranges from 9.28 eV for $C_6H_6$ to 7.16 eV for $C_{14}H_{10}$ and 7.43 eV for $C_{24}H_{12}$. Note that these values are much large as compared to the IE of alkali metal, whose maximal value is limited to 5.39 eV for Li [38]. With the substitution of Li atoms in $C_6H_6$, the VIE of $C_6Li_6$ is reduced to 4.48 eV, which is lower than the IE of Na (5.14 eV) [38]. Furthermore, the VIEs of $C_{10}Li_8$ and $C_{14}Li_{10}$ are even lower than the IE of K (4.34 eV) [38]. Although, the VIE of $C_{24}Li_{12}$ tends to be increased as compared to those of $C_{10}Li_8$ and $C_{14}Li_{10}$, but it is small enough as compared to those of Li and Na. Therefore, the VIEs of all lithiated analogues, being lower than those of alkali metals, suggest their superalkali behaviour. In order to explain the trend of VIE values, we have computed and listed sum of NBO charge values of Li atoms ($Q$) and average charge on Li ($q$) in Table 1. As mentioned earlier, traditional superalkalis employ a central (electronegative) atom attached with electropositive metal ligands. Lower IE of these superalkali species results due to increase in the charge delocalization over electropositive ligands. Analogously, Li atoms can be treated as ligands attached to carbon ring system as centre in these organic superalkalis. In case of $C_6Li_6$, net NBO charge on Li atoms is 3.72$e$ with an average value of 0.62$e$ per Li atom. For $C_{10}Li_8$ and $C_{14}Li_{10}$, $Q$ increases to 4.80$e$ and 6.10$e$, respectively, although $q$ does not change significantly. These causes to reduce VIE of these species relative to $C_6Li_6$. On the contrary, $Q$ value of $C_{24}Li_{12}$ is decreased to 5.88$e$ resulting in the increase in its VIE (see Table 1).

These organic superalkalis possess closed-shell planar structures, like their hydrocarbon analogues. It seems, therefore, interesting to compare the molecular orbitals of these superalkalis with those of parent hydrocarbons. In Fig. 2, we have plotted the highest occupied molecular orbital (HOMO) and delocalized molecular orbital of $C_6H_6$, $C_{10}H_8$, $C_{24}H_{12}$ and compared with those of $C_6Li_6$, $C_{10}Li_8$, $C_{24}Li_{12}$, respectively. The HOMO of $C_6H_6$ consists of delocalized electron cloud over C=C bond whereas HOMO-4 shows complete delocalization of electron over the whole ring. In the HOMO of $C_6Li_6$, the delocalization of electron cloud over C=C bond is reduced due to substitution of Li atoms. Table 1 lists the % contribution of atomic orbitals in the HOMOs of $C_6Li_6$, $C_{10}Li_8$, $C_{14}Li_{10}$ and $C_{24}Li_{12}$. One can note that the HOMO of $C_6Li_6$ is composed of 48% of Li and 52% of C atomic orbitals.

However, HOMO-7 of $C_6Li_6$ is completely delocalized, like HOMO-4 of $C_6H_6$. In case of $C_{10}Li_8$ and $C_{14}Li_{10}$, the contribution of Li atoms is 48% and 16%, respectively. On the contrary, the HOMO of $C_{24}Li_{12}$ is contributed by only C atoms (see Table 1), which resembles to that of $C_{24}H_{12}$ (see Fig. 2). Nevertheless, HOMO-12 of $C_{10}Li_8$ and HOMO-27 of $C_{24}Li_{12}$ possess delocalized π-electron cloud, corresponding to HOMO-7 of $C_{10}H_8$ and HOMO-23 of $C_{24}H_{12}$, respectively.

The complete delocalization of electrons is a distinct feature of aromatic ring systems. Therefore, we have quantified and analyzed the aromaticity of lithiated analogues and compared with corresponding hydrocarbons. In Table 2, we have listed NICS and $NICS_{zz}$ at the ring centre as well as HOMA and PDI values. $NICS_{zz}$, a tensor component of NICS along the direction perpendicular to the plane of ring, measures the contribution of π-electrons in the ring current. The positive value of NICS indicates anti-aromaticity whereas negative values confirm aromaticity. For instance, NICS and $NICS_{zz}$ values of $C_6H_6$ are -8.05 ppm and -14.49 ppm, which indicate its aromatic nature. This is further supported by its HOMA index (0.987) and PDI value (0.104). Although $NICS_{zz}$ value of $C_6Li_6$ is -25.11 ppm, its NICS value is only -3.38 ppm. This is due to contribution of σ-electrons from substituted from Li atoms, which supports paratropic ring currents and reduces the aromaticity significantly. This is also supported by smaller HOMA index (0.730) and PDI (0.091) of $C_6Li_6$. $C_{10}H_8$ contains two fused rings R1 and R2, which are as aromatic as $C_6H_6$ due to approximately similar NICS and $NICS_{zz}$ values, although HOMA and PDI values suggest the lower degree of aromaticity of $C_{10}H_8$ as compared to $C_6H_6$. The aromaticity of $C_{10}Li_8$ is slightly lower than that of its hydrocarbon analogue as suggested by NICS, $NICS_{zz}$, as well as PDI values but not by HOMA index, which suggests very poor aromaticity. The aromaticity of central ring of $C_{14}H_{10}$ (R2) is larger than those of accompanying rings (R1 and R3) but that of $C_{24}H_{12}$ (R4) is very less aromatic or not aromatic at all. This is supported by all aromaticity measures listed in Table 2. Likewise, the aromaticity of $C_{14}Li_{10}$ and $C_{24}Li_{12}$ is smaller than their hydrocarbon analogues as reflected by NICS and PDI values. The decrease in aromaticity can be expected due to Li-substitution. This feature has already been in some mono-substituted benzene [3, 4]. Therefore, our calculated NICS and PDI values clearly suggest that all lithiated analogues are indeed aromatic however their degree of aromaticity is lower than corresponding hydrocarbons.

We have already established that the IEs of $C_6Li_6$ and other lithiated analogues of higher hydrocarbons become lower than that of Li atom, just as those of superalkalis. For instance, the VIE of $C_6Li_6$ 4.48 eV is comparable to the IE of $FLi_2$ (4.20 eV) but smaller than that of $OLi_3$ (3.85 eV) superalkalis computed at the same level. Therefore, it seems interesting to study the interaction of $C_6Li_6$ with a superhalogen [39]. It has already been shown [12-15] that superalkalis interact with superhalogens and form ionic complexes, known as supersalts. We have studied the interaction of $C_6Li_6$ with $BF_4$ superhalogen and compared it with that of $OLi_3$ superalkali [8]. This interaction leads to formation of $C_6Li_6$-$BF_4$ and $OLi_3$-$BF_4$ complexes as displayed in Fig. 3. In both complexes, bonding between two units takes place via two Li atoms as shown. The binding energy of $OLi_3$-$BF_4$ is 8.55 eV in which the charge -0.91$e$ is transferred from $OLi_3$ to $BF_4$. Therefore, $OLi_3$-$BF_4$ is an ionic complex, which can be written as $(OLi_3)^+(BF_4)^-$. Likewise, the charge transfer from $C_6Li_6$ to $BF_4$ is -0.87$e$ such that the binding energy of $C_6Li_6$-$BF_4$ is reduced only slightly (7.80 eV). Nevertheless, this complex can be also considered as $(C_6Li_6)^+(BF_4)^-$. Thus, $C_6Li_6$ interacts with superhalogen more or less similarly as a typical superalkali does.

Borazine ($B_3N_3H_6$) is an inorganic analogue of $C_6H_6$. Lithiated borazine ($B_3N_3Li_6$) has also been reported to be star-like structure [40]. Our recent study [41] shows that the VIE of $B_3N_3Li_6$ is 4.25 eV, which is comparable to that of $C_6Li_6$. Therefore, it seems interesting to compare the bonding feature of $C_6Li_6$ and $B_3N_3Li_6$. We have calculated molecular graph of both $C_6Li_6$ and $B_3N_3Li_6$ using quantum theory of atoms in molecule (QTAIM) method [42, 43] as displayed in Fig. 4. In the framework of QTAIM, the bonding between two atoms is characterized by a bond critical point (BCP), shown by green points in Fig. 4. In the molecular graph of $C_6Li_6$, we obtain two BCPs for each Li atom, which correspond to C-Li bonds. Therefore, $C_6Li_6$ is indeed a star-like structure in which Li atoms interact with two neighbouring C atoms of the ring. Likewise, we find two BCPs for each Li atoms in $B_3N_3Li_6$. In contrast to $C_6Li_6$, both BCPs in $B_3N_3Li_6$ correspond to N-Li bonds, no B-Li bonds. Thus, there exists no B-Li bond in $B_3N_3Li_6$ according to QTAIM analysis. Consequently, $B_3N_3Li_6$ is not strictly star-like rather it appears as a fan-like structure. This is evidently due to large electronegativity difference between B and N atoms in $B_3N_3Li_6$, which is not the case for $C_6Li_6$.

## 4. Conclusions

We have performed density functional investigations on Li-substituted at all H atoms of aromatic hydrocarbons such as benzene ($C_6H_6$), naphthalene ($C_{10}H_8$), anthracene ($C_{14}H_{10}$) and coronene ($C_{24}H_{12}$). Our study leads to conclude following:

a) Unlike corresponding hydrocarbons, the VIEs of all lithiated species are lower than the IE of Li atom, just like those of superalkalis. However, they all possess closed shell structures, unlike superalkalis.

b) Like corresponding hydrocarbons, all lithiated species are planar and aromatic. However, their aromaticity is reduced due to Li-substitution as compared to hydrocarbons.

c) The interaction of $C_6Li_6$ with $BF_4$ superhalogen is similar to that of $OLi_3$ superalkali. This may suggest that these lithiated species behave as superalkalis.

d) $C_6Li_6$ possesses star-like structure, unlike its inorganic analogue $B_3N_3Li_6$, which is found to be fan-like structure rather than star-like as proposed earlier.

**Acknowledgement**

Dr. A. K. Srivastava acknowledges Prof. N. Misra, Department of Physics, University of Lucknow for providing computational facilities and helpful suggestions.

Table 1. Calculated bond lengths and vertical ionization energy of lithiated species along with their hydrocarbon analogues. NBO charges and HOMO-composition are also listed,

| Systems (Sym.) | d(C−C) (Å) | d(C−Li) (Å) | VIE (eV) | NBO charge on Li | Composition of HOMO | Hydrocarbon analogues | | |
|---|---|---|---|---|---|---|---|---|
| | | | | | | Systems | d(C−C) (Å) | VIE (eV) |
| $C_6Li_6$ ($C_{2v}$) | 1.420 | 1.914 | 4.48 | $Q = 3.72$ $q = 0.62$ | C = 52% Li = 48% | $C_6H_6$ | 1.395 | 9.28 |
| $C_{10}Li_8$ ($D_{2h}$) | 1.390−1.463 | 1.896−1.962 | 4.29 | $Q = 4.80$ $q = 0.60$ | C = 34% Li = 48% | $C_{10}H_8$ | 1.374−1.420 | 7.97 |
| $C_{14}Li_{10}$ ($C_{2h}$) | 1.382−1.468 | 1.919−1.963 | 4.24 | $Q = 6.10$ $q = 0.61$ | C = 16% Li = 36% | $C_{14}H_{10}$ | 1.367−1.443 | 7.16 |
| $C_{24}Li_{12}$ ($C_6$) | 1.383−1.450 | 1.904−1.953 | 4.50 | $Q = 5.88$ $q = 0.49$ | C = 74% Li = 0% | $C_{24}H_{12}$ | 1.370−1.426 | 7.43 |

Table 2. Various aromaticity descriptors of of lithiated species along with their hydrocarbon analogues. Please refer to Fig. 1 for labelling of rings.

| Para-meter | X | $C_6X_6$ | $C_{10}X_8$ | | $C_{14}X_{10}$ | | | $C_{24}X_{12}$ | | | | | | |
|---|---|---|---|---|---|---|---|---|---|---|---|---|---|---|
| | | | R1 | R2 | R1 | R2 | R3 | R1 | R2 | R3 | R4 | R5 | R6 | R7 |
| NICS (ppm) | H | -8.05 | -8.39 | -8.39 | -7.28 | -11.05 | -7.28 | -9.60 | -9.72 | -9.72 | -0.22 | -9.72 | -9.72 | -9.60 |
| | Li | -3.38 | -7.74 | -7.74 | -5.21 | -11.59 | -5.21 | -7.84 | -7.84 | -7.69 | 0.35 | -7.69 | -7.84 | -7.84 |
| $NICS_{zz}$ (ppm) | H | -14.49 | -13.29 | -13.29 | -9.68 | -19.02 | -9.68 | -13.41 | -13.68 | -13.68 | 16.94 | -13.68 | -13.68 | -13.41 |
| | Li | -25.11 | -25.22 | -25.22 | -18.92 | -23.79 | -18.92 | -15.59 | -15.59 | -15.11 | -19.31 | -15.11 | -15.59 | -15.59 |
| HOMA | H | 0.987 | 0.783 | 0.783 | 0.629 | 0.720 | 0.629 | 0.736 | 0.736 | 0.736 | 0.627 | 0.736 | 0.736 | 0.736 |
| | Li | 0.730 | 0.295 | 0.295 | 0.016 | 0.294 | 0.016 | 0.402 | 0.402 | 0.402 | 0.378 | 0.402 | 0.402 | 0.402 |
| PDI | H | 0.104 | 0.076 | 0.076 | 0.067 | 0.066 | 0.067 | 0.054 | 0.054 | 0.054 | 0.033 | 0.054 | 0.054 | 0.054 |
| | Li | 0.091 | 0.068 | 0.068 | 0.060 | 0.061 | 0.060 | 0.051 | 0.051 | 0.051 | 0.031 | 0.051 | 0.051 | 0.051 |

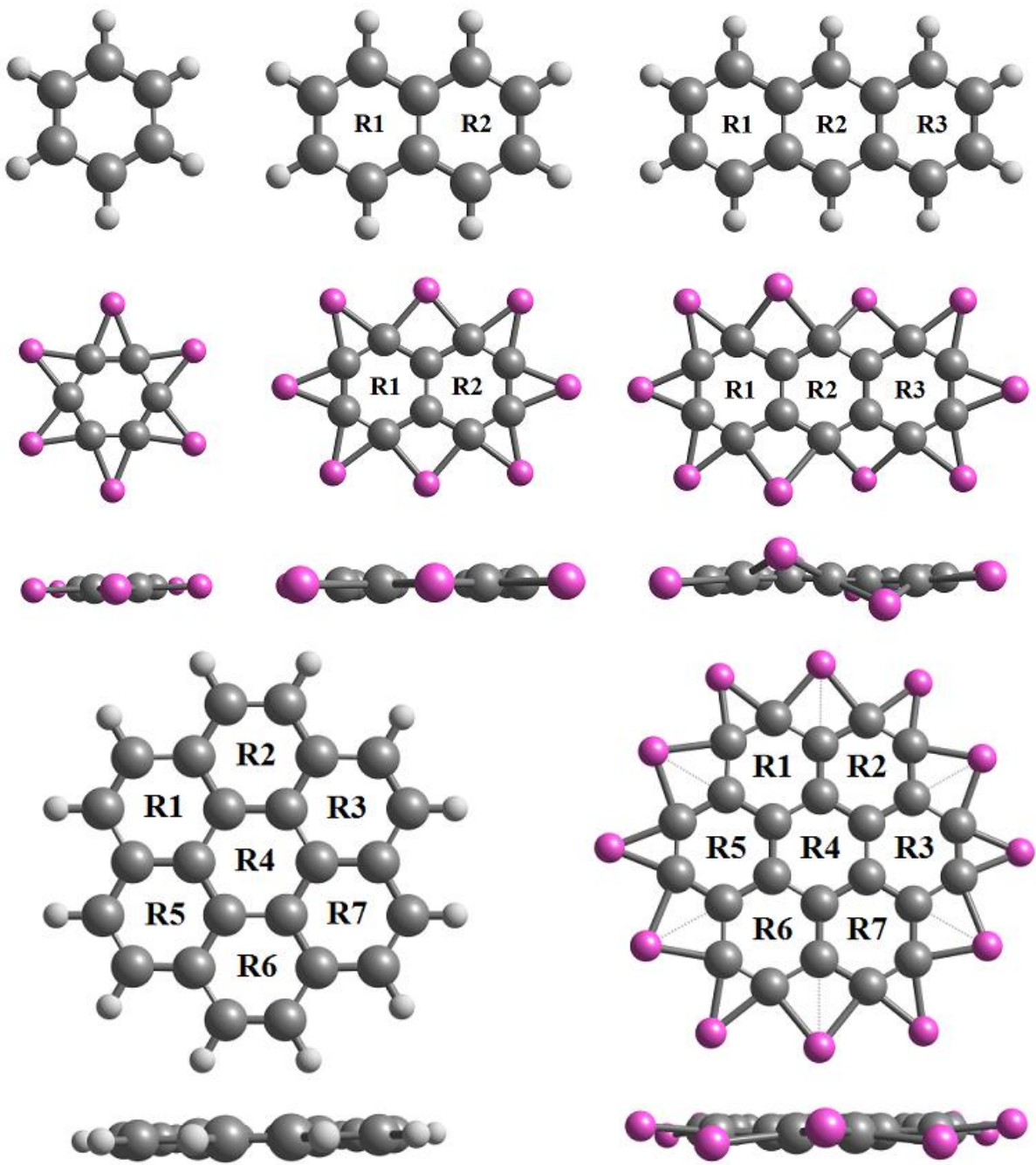

Fig. 1. Equilibrium structures of benzene ($C_6H_6$), naphthalene ($C_{10}H_8$), anthracene ($C_{14}H_{10}$) and coronene ($C_{24}H_{12}$) along with their lithiated analogues obtained at B3LYP level.

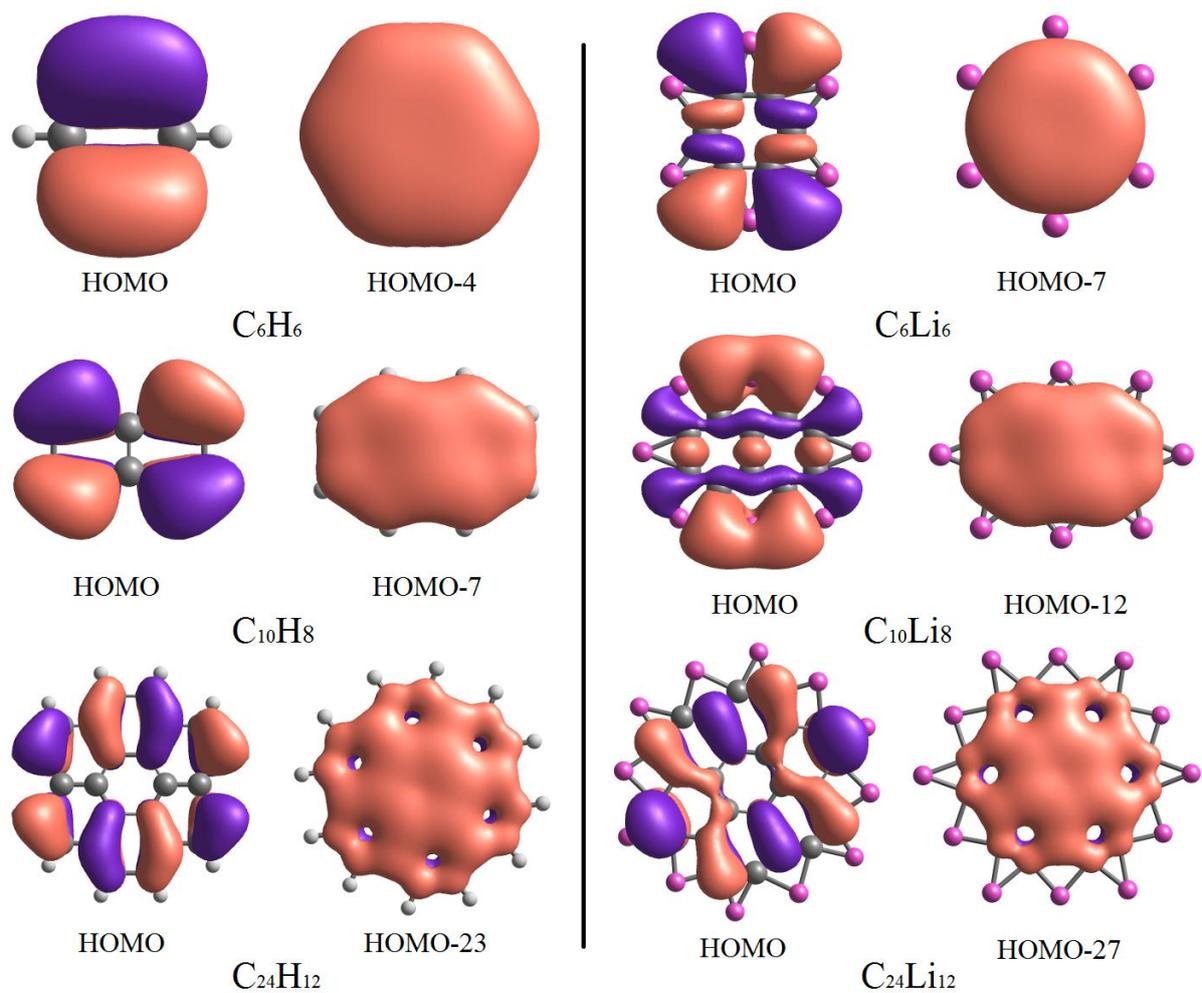

Fig. 2. Selected molecular orbital surfaces including the highest occupied molecular orbital (HOMO) of the systems considered in this study with an isovalue of 0.02 a.u.

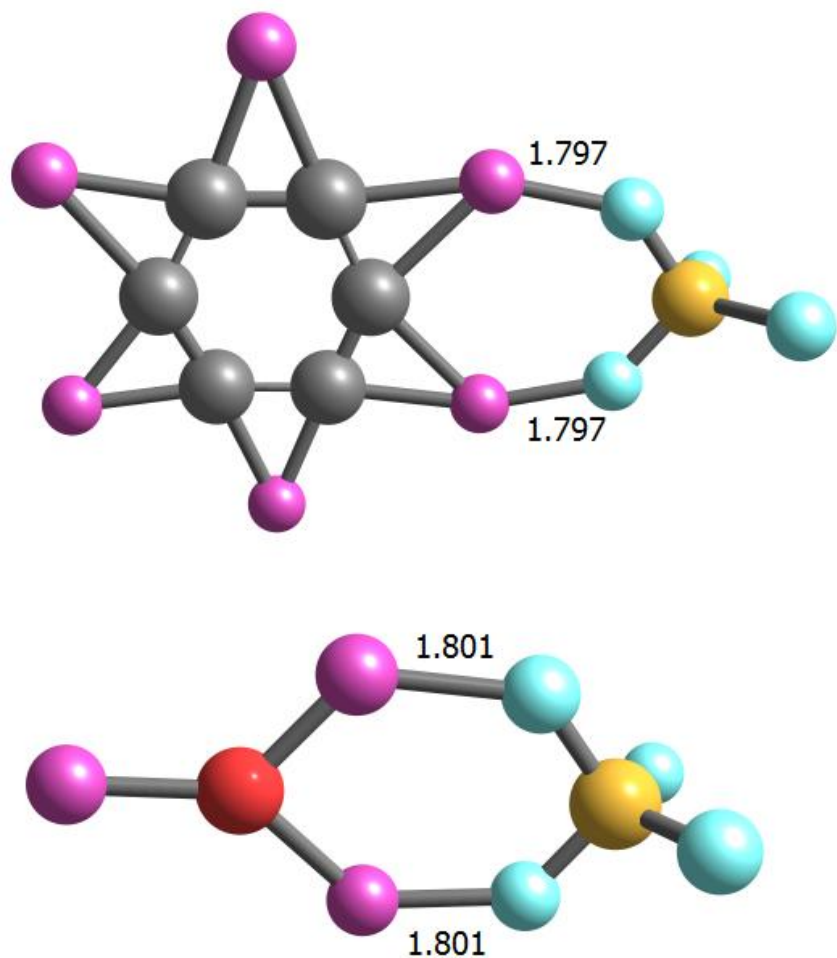

Fig. 3. Equilibrium structure of $C_6Li_6$-$BF_4$ and $OLi_3$-$BF_4$ complexes obtained at B3LYP/6-311+G(d) level. The interaction bond-lengths (in Å) are also displayed.

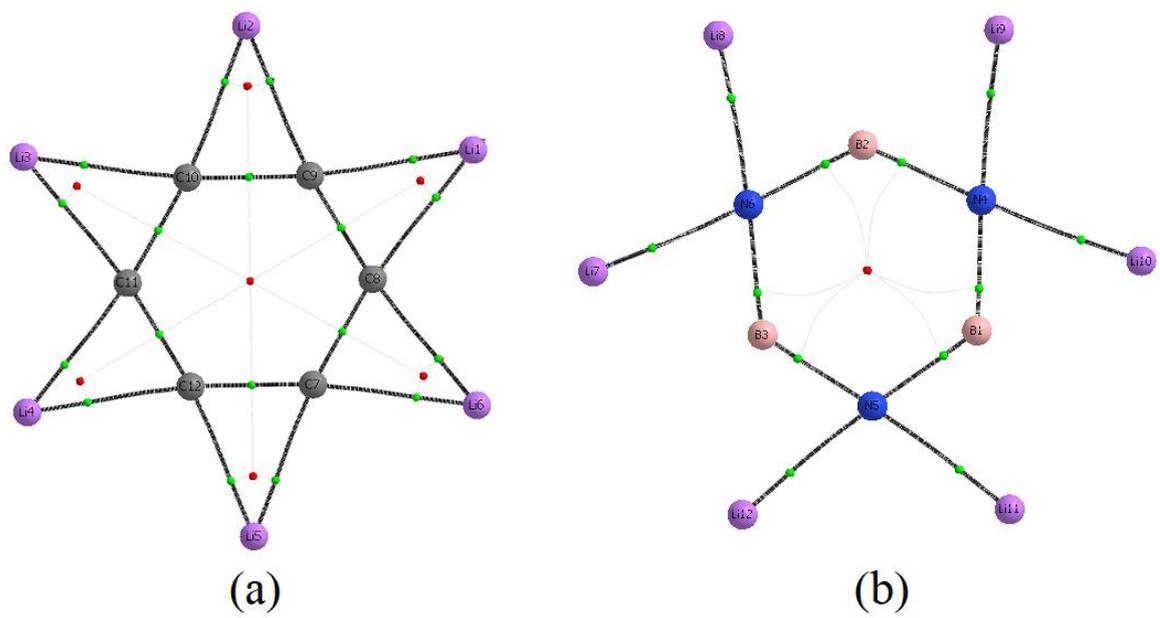

Fig. 4. Molecular graphs of $C_6Li_6$ (a) and $B_3N_3Li_6$ (b) computed by QTAIM method. BCPs are shown by green points.